\documentclass[aps,prb,twocolumn,showpacs,groupedaddress,floatfix]{revtex4}
\usepackage{graphicx}
\begin{document}


\title{Influence of interface structure on electronic properties \\
and Schottky barriers in Fe/GaAs magnetic junctions}

\author{D. O. Demchenko and Amy Y. Liu}
\affiliation{Department of Physics, Georgetown University, 
  Washington, DC 20057-0995, U.S.A.}
\date{\today}

\begin{abstract}
The electronic and magnetic properties of Fe/GaAs(001) magnetic 
junctions are investigated using first-principles density-functional 
calculations.  Abrupt and intermixed interfaces are considered,
and the dependence of charge transfer, magnetization
profiles, Schottky barrier heights, and spin polarization of densities 
of states on interface structure is studied.   With As-termination,
an abrupt interface with Fe is favored, while Ga-terminated
GaAs favors the formation of an intermixed layer with Fe.    The
Schottky barrier heights are particularly sensitive to 
the abruptness of the interface.  
A significant density of states in the semiconducting gap arises from 
metal interface states.
These spin-dependent interface
states lead to a significant minority spin polarization of the 
density of states at the Fermi level that persists well into the 
semiconductor,
providing a channel for the tunneling of minority spins through
the Schottky barrier.   These interface-induced gap
states and their dependence on atomic structure at the interface are
discussed 
in connection with potential spin-injection
applications.

\end{abstract}


\maketitle
\section{Introduction}
The problem of spin injection has been a subject of intense study since the
proposal of the electronic analog of the electro-optic modulator. \cite{datta}
Since then the entire field of spintronics has been developed. The main idea
lies in the possibility of controlling the spin of charge carriers, thereby
adding an additional degree of freedom to existing semiconductor-based
electronics. 
A number of novel electronics devices based on this idea have been proposed, 
such as reprogrammable logic devices, spin valves, spin-injection
diodes, and devices utilizing giant magnetoresistance 
(see, for instance, Ref. \onlinecite{spintronics}).
A major challenge in the field has been the creation of spin-polarized 
currents in nonmagnetic semiconductors.  One  approach is the use of
ferromagnetic contacts as spin sources.  A spin polarization of the
current is expected from the different conductivities resulting
from the different densities of states for spin-up and spin-down
electrons in the ferromagnet.
Significant progress in molecular beam epitaxy  has allowed
growth of high quality, virtually defect-free junctions between 
magnetic materials and semiconductor substrates, and 
films of ferromagnetic metals such as Fe or Co grown epitaxially on
semiconductor structures are promising candidates for spin injection.

Zhu and co-authors \cite{Zhu} have  
demonstrated efficiencies of 2\% for 
injection of spin-polarized electrons from a metal into a semiconductor 
for a GaAs/(In,Ga)As light emitting diode (LED) covered with Fe. 
Moreover, Hanbicki {\it et al}. \cite{Hanbicki}
have managed to achieve a spin injection efficiency of 30\%.  
In the latter case, an Fe film grown on an AlGaAs/GaAs quantum well LED 
structure was used. 
In both cases it was suggested that the spin injection arises from
tunneling of spin-polarized electrons from the metal into the semiconductor 
across the Schottky barrier formed at the interface. 
Such a tunneling process is believed to be responsible for 
the spin injection since it is not affected by the conductivity 
mismatch \cite{Rashba} between the metal and semiconductor that 
severely limits the spin-injection efficiency in the diffusive transport 
regime.  
A remarkable consequence of such a 
mechanism is that it is independent of temperature.  In fact, in 
Refs. \onlinecite{Zhu} and \onlinecite{Hanbicki},
nearly constant tunneling efficiencies  of 2\% and 30\% were observed for a 
range of temperatures from 2 K to 300 K and  from 90 K to 240 K, respectively. 
Recently spin injection efficiencies of 13\% have been 
reported at 5 K across a Fe/GaAs(110) interface, \cite{Jonker_new} and
6\% across a Fe/Al$_x$Ga$_{1-x}$As/GaAs Schottky contact at 295 K. \cite{Minnesota_new} 
At the moment, the room temperature record of 32\% is held by CoFe/MgO 
injectors grown on $p$-GaAs(100) substrates. \cite{Parkin}

The electronic structure of the metal/semiconductor
interface plays an important role in spin-dependent transport properties
across such junctions.  
For the case of the Fe/GaAs(001) interface, a number of first-principles
studies have been carried out.  Green's function
methods have been used to study the electronic structure, charge transfer,
and spin polarization in Fe/GaAs/Fe(001) tunnel junctions in which the 
interface geometry is atomically abrupt and 
ideal.\cite{Butler,Dederichs1,Vlaic}
It was noted that the
calculated magnetic properties at the interface are sensitive to the
interface structure, indicating that structural relaxations could be
important.\cite{Dederichs1}  Other studies have focused on the initial stages 
of growth of Fe on GaAs and have considered how 
structural relaxation and intermixing of metal and semiconductor atoms affect 
magnetic properties of thin films of Fe on GaAs. \cite{Erwin,Mirbt,Kim}
Still lacking is an understanding of how details of the
atomic arrangement at the interface affect properties directly relevant to spin
injection, such as Schottky barrier heights  and the nature of 
the interface-induced states in the semiconductor gap through which
spins may tunnel. 
In this paper, we address this by considering structural models
for the (001) interface between Fe and GaAs that allow for different
degrees of intermixing and relaxation. The 
effects of interface structure on potential, charge, and magnetization
profiles, Schottky barrier heights, spin-polarized densities of states,
and interface-induced gap states are investigated. 
The results are discussed in connection 
with potential spin-injection applications.

The paper is organized in the following way. 
Computational details are given in Section II. 
In Section III we describe and discuss the structures considered 
and the resulting 
structural, electronic, and magnetic properties. 
Conclusions are presented in Section IV.

\section{Computational Details}

The bulk lattice constants of bcc Fe (2.866 \AA) and zinc-blende GaAs (5.654 \AA)
differ by almost exactly a factor of two. Therefore, an interface consisting 
of a (001) Fe slab placed on the (001) surface of GaAs has a lattice mismatch 
of 1.36\%.
The close lattice match
helps keep the concentration of defects at such interfaces relatively low.
In this work, interfaces were modeled using 
supercells consisting of nine layers of Fe 
and nine layers of semiconductor atoms (five As and four Ga, or vice versa, depending
on the semiconductor surface termination.) 
This is sufficient to ensure that the adjacent
interfaces do not interact with each other, as confirmed by both total energy
calculations and macroscopic averages of the electrostatic potential and charge. 

The calculations were performed with the VASP program, \cite{vasp} 
an {\it ab initio} density functional 
code that uses the planewave pseudopotential method. The generalized gradient 
approximation 
was used to treat the exchange-correlation part of the electron-electron 
interaction.\cite{GGA1,GGA2} All calculations employed ultrasoft pseudopotentials \cite{Vanderbilt} and planewave basis sets with a kinetic energy cut off of 
370 eV. 
Monkhorst-Pack meshes\cite{Monkhorst-Pack}
of $8\times 8\times 4$ {\bf k}-points were used to sample the
Brillouin zone, and Gaussian smearing of electronic states 
($\sigma = 0.2$ eV) was used to achieve faster convergence of
Brillouin zone sums with respect to the number of {\bf k}-points.

\section{Results}

\subsection{Structure}

\begin{figure}[tbfh]
\includegraphics[width=3.4 in,height=2.0 in]{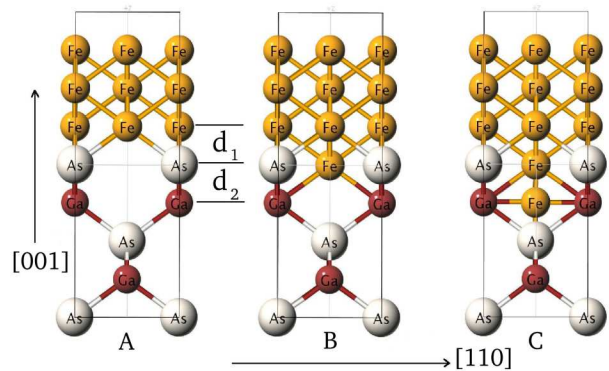}
\caption{Three models for the structure of the $1\times1$ interface
between As-terminated GaAs(001) and Fe(001) surfaces. 
Model A is an abrupt interface, model B  has one layer containing
both metal and semiconductor atoms, and model C has two intermixed
layers. 
\label{fig: ABC}}
\end{figure}

The Fe/GaAs interface structures considered in this work are shown in Fig. \ref{fig: ABC}. 
While the bare GaAs(001) surface reconstructs to form dimers on the surface, 
\cite{dimers}
recent calculations  suggest that these dimers become unstable upon 
adsorption of Fe.\cite{Erwin}    Therefore, 
in this work, we consider 1$\times$1 interfaces only.  We assume 
the Fe(001) and GaAs(001) slabs are aligned so that in the first complete 
metal layer, sites that would have been occupied by semiconductor atoms in the 
absence of the interface are now occupied by metal atoms. An abrupt 
interface like this with no intermixing of metal and semiconductor
atoms will be referred to as model A (as in Ref. \onlinecite{Erwin}). In model B, the bcc metal structure is 
partially continued into the first semiconductor layer so that metal atoms occupy 
sites that are normally empty in the first layer of the semiconductor. 
In model C, metal atoms  occupy interstitial sites in the second semiconductor layer as well.      
Both As- and Ga-terminated interfaces can
be grown experimentally, and therefore, both are considered here. 
We find that for the supercells of the size considered in this work,
model C is never energetically favorable. This agrees with 
Ref. \onlinecite{Erwin},
where it was found that model C has lower energy only for coverages of Fe
not exceeding two monolayers. Therefore we focus on models A and B.

The comparison of total energies of models A and B is not physically meaningful 
since the interfaces have different numbers of atoms. The formation energies 
of these models are, on the other hand, physically comparable and given by 
\begin{equation}
E_{form}=E_t-\sum_{i}N_i\mu_i .
\label{eq: form_energy}
\end{equation}
Here, $E_t$ is the total energy of the supercell, $N_i$ is the number of atoms 
of the type $i$ in the cell, and $\mu_i$ is the chemical potential of the $i$-th
atom. Thus, all energies from the model B calculations were adjusted by 
twice the value of the chemical potential for Fe, 
since each supercell contains two equivalent interfaces. 
Test calculations of the formation energy 
using supercells with one A interface and one B interface 
confirm that the supercells are large enough to ensure that the interaction
between interfaces in negligible. 

Since electronic and magnetic properties may differ significantly as a 
result of small changes in structural parameters, we have relaxed all 
the interface models 
considered. In our calculations the interfaces were relaxed with respect to two 
parameters: the distance between the adjacent Fe and As (or Ga) layers, $d_1$, 
and the distance between the first two layers in the GaAs slab, $d_2$ 
(see Fig. \ref{fig: ABC}). The in-plane lattice constant was fixed at the bulk 
GaAs value, $a_0 = 5.654$ \AA. 
The results of the relaxation 
are presented in Table \ref{tab: relaxation}.
\begin{table}[t]
\caption{Interlayer relaxations, in units of the GaAs lattice constant $a_0$.
Here, $\Delta d_1$ is the difference between the relaxed and ideal ($0.25a_0$)
separation  of adjacent Fe and As (Ga) planes at the interface, and 
$\Delta d_2$ is the change in the distance between the first two 
planes in the GaAs slab. 
\label{tab: relaxation}}
 \begin{ruledtabular}
\begin{tabular}{ccccc}
  & \multicolumn{2}{c}{As-terminated} & \multicolumn{2}{c}{Ga-terminated} \\
\hline
  & $\Delta d_1$ & $\Delta d_2$ & $\Delta d_1$ & $\Delta d_2$ \\
\hline
Model A & 0.0 & 0.025 & 0.022 & 0.017 \\
Model B & 0.017 & 0.068 & 0.018 & 0.058 
\end{tabular}
 \end{ruledtabular}
\end{table}
For As-terminated model A, the distance between As and Fe planes remains at 
the ideal value of
0.25$a_0$, while the separation between the first plane of As and the
adjacent plane of Ga is stretched
by 10\%. With Ga-termination, both $d_1$ and $d_2$ increase by
similar amounts in model A. In model B, which has the intermixed layer,
the first two planes of the semiconductor slab are
repelled considerably farther apart, regardless of the termination. 
It was previously pointed out 
that lowering the concentration of atoms in the interface region is 
energetically favorable in model B, since the electrons from the extra Fe 
atoms  in the intermixed layer
fill antibonding orbitals and weaken the interface bonding.\cite{Erwin} 

The results of the formation energy calculations for the ideal and relaxed interfaces are
summarized in Table \ref{tab: energies}. 
\begin{table}[b]
\caption{Formation energy differences, in eV per  
1$\times$1 interface unit cell.  With As-termination,  the 
abrupt interface of model A is favored, while with Ga-termination,
the intermixed interface  of model B is preferred. 
\label{tab: energies}}
 \begin{ruledtabular}
\begin{tabular}{ccc}
  & As-terminated & Ga-terminated \\
\hline
  & Ideal - Relaxed & Ideal - Relaxed \\
\hline
Model A & 0.046 & 0.059  \\
Model B & 0.282 & 0.291   \\
\hline
& Model A - Model B & Model A - Model B \\
\hline
Ideal &   -0.400  &  0.318   \\
Relaxed & -0.164  &  0.551
\end{tabular}
 \end{ruledtabular}
\end{table}
For both ideal and relaxed geometries, model A is energetically favored for
As-terminated interfaces, and model B is favored for Ga-terminated
interfaces. 
This results from an interplay between optimization of the coordination of the
metal  and semiconductor atoms and the relative strengths of the
metal-cation and metal-anion bonds.  Because of stronger $pd$ hybridization,
the Fe-As bond is more stable than the Fe-Ga bond. \cite{Mirbt}
For model A, the interface Fe site is six-fold
coordinated, with four Fe neighbors in the second metal layer and two 
semiconductor neighbors in the first semiconductor layer.  While  putting 
Fe atoms in
interstitial sites in the first semiconductor layer fully coordinates
the interface Fe sites in model B, it also significantly weakens the bonding
between the first two semiconductor layers, which become overcoordinated.
At the As-terminated interface, the strong Fe-As bonds compensate for
the undercoordination of the Fe sites, making model A favorable, while 
at the Ga-terminated interface, the weak interaction between
Fe and Ga favors full coordination of Fe interface sites, as in model B. 

\subsection{Schottky Barriers and Electronic Structure}
\begin{figure}[tbf]
\includegraphics[width=3.4 in,height=4.0 in]{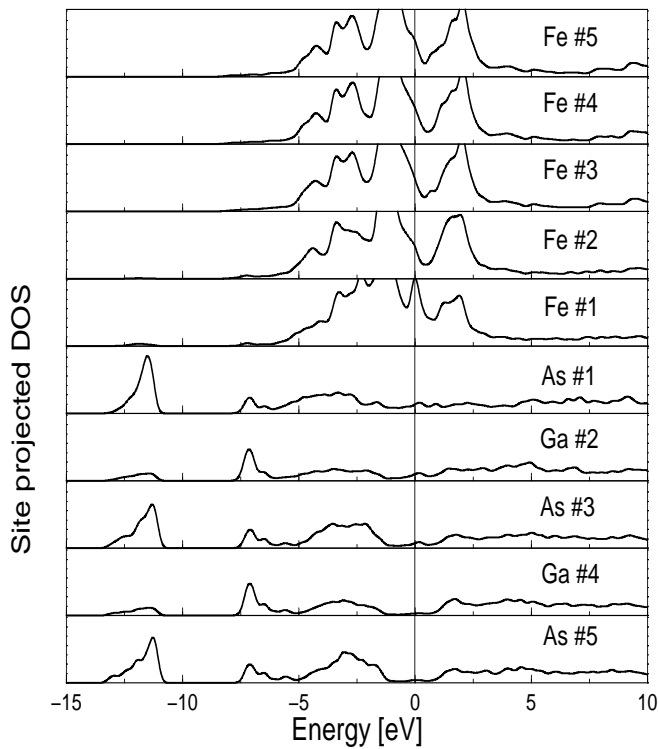}
\caption{Site-projected densities of states  for atoms in different
atomic layers of the relaxed GaAs/Fe(001)  supercell
with As-terminated model A interfaces.
The bottom panel represents the semiconductor layer farthest from the interface
(As \#5),  and the top panel represents 
the most bulk-like layer of Fe (Fe \#5).  The vertical line at zero energy 
indicates the Fermi level. In each panel, the vertical scale
runs from 0 to 1.8 states/eV/atom. 
\label{fig: DOS}}
\end{figure}

Figure \ref{fig: DOS} shows the calculated site-projected 
densities of states (DOS) for atoms in different layers in
the relaxed As-terminated model A. The general features are similar
for all the structural models considered. The DOS in the most bulk-like 
layers of the supercell closely resembles the DOS of bulk Fe or GaAs. 
Near the interface, a peak develops in the DOS at the Fermi level.
The peak is largest in the first Fe layer and decreases into both the 
GaAs and Fe slabs. In the first few GaAs layers, 
states spread throughout the entire band gap of bulk GaAs, though
the gap is practically recovered in the layer farthest from the interface.
Bardeen \cite{Bardeen} estimated that surface
states with density $\sim 10^{13}$ eV$^{-1}$ cm$^{-2}$ 
would effectively pin the Fermi level close to the 
charge neutrality level in the semiconductor gap. 
Our calculations for the relaxed As-terminated model A, for example, 
yield values of DOS at the Fermi level
of $\sim 2 \times 10^{15}$ eV$^{-1}$ cm$^{-2}$ in the first Fe layer, and 
$\sim 1 \times 10^{14}$ eV$^{-1}$ cm$^{-2}$ 
in the first two layers of the GaAs slab. 
These may be sufficient to pin the Fermi level, which consequently affects 
the Schottky barrier height (SBH).   We will return to examine
the nature of these mid-gap states, their spin polarization, and
their role in spin tunneling 
in section D. 

As mentioned, Schottky barriers may be a necessary mechanism 
for overcoming the conductivity mismatch for the injection of spin from a 
ferromagnetic metal into a semiconductor. 
Schottky barrier heights can be evaluated from first principles using a 
macroscopic average method combined with the supercell 
approach. \cite{Baldereschi} 
Two necessary
conditions are: 1) the supercell contains two equivalent interfaces, which 
eliminates any electric fields that might be present due to unbalanced
charges, and 2) the supercell is sufficiently large so the bulk charge 
and potential properties are recovered in the most bulk-like layers of the 
supercell. In such a way an isolated interface is accurately
modeled using  a supercell. 

The calculation of the $p$-type Schottky barrier height, $\phi_p$, 
is split into two parts: 
\begin{equation}
\phi_p = \Delta E_v + \Delta V.
\label{eq: schottky}
\end{equation}
The band structure term, 
\begin{equation}
\Delta E_v = E_F - E_v,
\label{eq: band_term}
\end{equation}
is the difference between the Fermi level $E_F$ in the metal and the valence
band edge in the semiconductor $E_v$, where each is measured with respect to 
the average electrostatic potential in the corresponding bulk material. 
The band structure term is calculated from separate
bulk calculations for the two constituents of the Schottky contact. This term 
implicitly includes all quantum mechanical effects as well as the exchange-correlation
part of the potential. 

\begin{figure}[tbf]
\includegraphics[width=3.0 in,height=2.70 in]{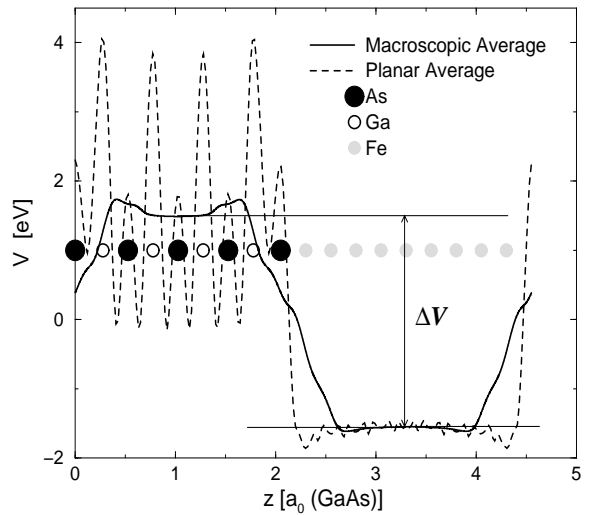}
\caption{Planar and macroscopic averages of the electrostatic
potential calculated for the relaxed As-terminated  model A supercell. 
The potential is plotted as a function of $z$, the position  of
atomic layers along the axis of the supercell. 
The potential line-up $\Delta V$ needed for the Schottky barrier
height calculation is given by the difference between the two plateaus 
in the macroscopically averaged potential.
\label{fig: pot_As}}
\end{figure}

The other contribution to $\phi_p$ is the potential line-up across the 
interface $\Delta V$.  This potential line-up is related to
the dipole moment of the charge profile,\cite{Baldereschi} 
depends on the structure of the interface, and therefore cannot be 
calculated simply from two bulk calculations. 
It is the difference between the macroscopic averages of the electrostatic 
potential in two bulk-like regions of the supercell. 
As an example, Fig. \ref{fig: pot_As} shows the planar average over the $x-y$ plane 
and the macroscopic average of the electrostatic
potential computed for the superlattice with As-terminated relaxed model A
interfaces. 
The macroscopic average lacks the bulk-like
oscillations present in the planar average, 
and thereby allows one to extract the desired potential 
line-up $\Delta V$. 
The macroscopic average, however, does not provide reliable information 
about potential behavior in the close vicinity of the interface since 
the resolution of the method is limited to the lattice period, the
GaAs lattice constant $a_0$ in our case. 
There is a notable difference between the planar average and the macroscopic average in the 
vicinity of the interface. 

We have calculated the $p$-type SBH for the ideal and relaxed structures 
described in the previous section.
Since it is electrons that are injected from Fe into GaAs in spin injection 
experiments, the $n$-type SBH is of more interest. 
To obtain the $n$-type SBH from our calculations, we subtract the 
calculated $p$-type SBH from the experimental GaAs band gap of 1.52 eV. \cite{Aspnes}
The results are presented in Table \ref{tab: schottky}. 
\begin{table}[b]
\caption{ Calculated {\sl n}-type Schottky barrier heights (in eV) 
for ideal and relaxed geometries.  Interfaces with intermixing (model B)
have lower barrier heights than abrupt interfaces (model A).
\label{tab: schottky}}
 \begin{ruledtabular}
\begin{tabular}{ccccc}
  & \multicolumn{2}{c}{As-terminated} & \multicolumn{2}{c}{Ga-terminated} \\
\hline
  & Ideal & Relaxed & Ideal & Relaxed \\
\hline
Model A & 0.87 & 0.82 & 1.01 & 1.08 \\
Model B & 0.58 & 0.64 & 0.69 & 0.89 
\end{tabular}
 \end{ruledtabular}
\end{table}
The calculated SBH is sensitive to structural changes, and the differences 
between the SBH values in Table \ref{tab: schottky} show the magnitude of 
changes one can expect for different interface structures. 
Intermixing of metal and semiconductor atoms at the interface 
decreases the $n$-type Schottky barrier height significantly (by about
0.2-0.3 eV), while the effect of interlayer relaxation is generally weaker.
This level of sensitivity of SBH values to interface structure is
consistent with the observed dependence of measured SBHs 
on growth conditions. 
For Fe/GaAs interfaces fabricated by metal 
evaporation in ultrahigh vacuum, SBHs in the range of 0.72-0.75 eV have
been reported.\cite{experiment}  More recent experiments on As-terminated
atomically clean (001) interfaces grown by molecular beam epitaxy 
have yielded barriers
around 0.90-0.92 eV.\cite{experiment_Jonker1,experiment_Jonker2}
In the latter work, the samples had sufficiently low interface defect
concentrations that Fermi-level pinning by antisite defects was suppressed. 
\cite{experiment_Jonker1, experiment_Jonker2}   
Our calculated results for the energetically favored
geometries, i.e. As-terminated relaxed model A and
Ga-terminated relaxed model B,
are close to the SBH values obtained for these clean samples.
Overall, the present results compare more favorably to experiments than
earlier density-functional calculations of the $p$-type SBH at 
ideal Fe/GaAs (110) interfaces.\cite{Mark}

\subsection{Charge Distribution and Magnetic Moments}

\begin{figure}[t]
\includegraphics[width=3.0 in,height=2.50 in]{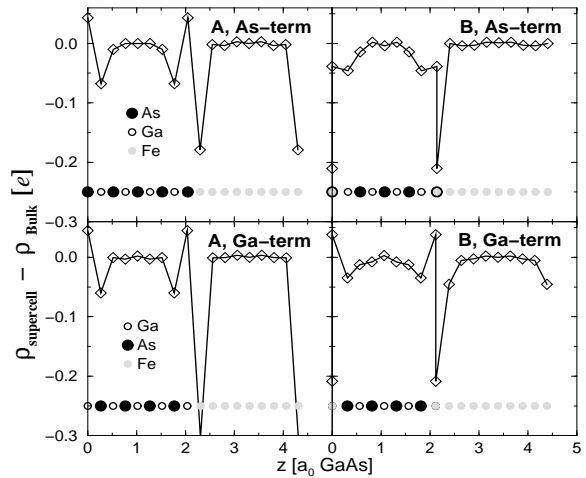}
\caption{Difference between the charge inside the atomic spheres for 
the supercell and bulk GaAs and Fe, plotted as a function of distance $z$
along the direction normal to the interface. The upper panels show results for 
relaxed As-terminated models A and B, and the lower panels show results for  
relaxed Ga-terminated models A and B.  
In the intermixed layer in model B,
the higher point represents the charge difference for the As 
or Ga atom and the lower point represents that for the Fe atom. 
\label{fig: WZ}}
\end{figure}

To investigate the redistribution of charge that  
gives rise to the Schottky barrier, we have computed the charge within 
Wigner-Seitz spheres centered on atomic sites.\cite{WS}
We have also compared the planar-averaged charge in the interface region to
the planar-averaged charge in the bulk-like regions and to the planar-averaged
superposition of atomic charges. 
The qualitative features of the
charge redistribution deduced from the sphere charges 
are consistent with those suggested by the planar-averaged charge.
Figure \ref{fig: WZ} shows the difference 
between the charge inside the spheres in the supercell and in the bulk. 
At the abrupt interface (model A), 
there is an evident transfer of charge from the 
interface Fe layer into the semiconducting slab, regardless of the 
semiconductor termination. In  Ref. \onlinecite{Dederichs1}, it was
found that for unrelaxed abrupt interfaces, the interface Fe loses more
charge to Ga neighbors than to As neighbors.  We find that this
difference is enhanced when the interlayer distances are allowed to relax.
Interface Fe atoms also lose significant charge to the interstitial region
where bond formation takes place.   The redistribution of
charge is limited to regions close to the interface. By the third
semiconductor layer, the charge within the spheres has nearly recovered
to bulk values.

In model B, Fe sites in the first full Fe layer are fully coordinated,
so the local charge around these sites is much closer to the bulk value than
in model A.  The charge redistribution takes place primarily within the
intermixed layer, where the Fe sites are undercoordinated and the
semiconductor sites are overcoordinated.  While the amount of
local charge lost by Fe sites in the intermixed layer is about the same
for both terminations ($\sim0.2e$), with Ga-termination, the Ga sites in the 
intermixed layer gain electrons while with As-termination, the As interface 
sites lose electrons, indicating a transfer of charge into interstitial
regions in the Fe-As intermixed layer. 

\begin{figure}[t]
\includegraphics[width=3.0 in,height=2.50 in]{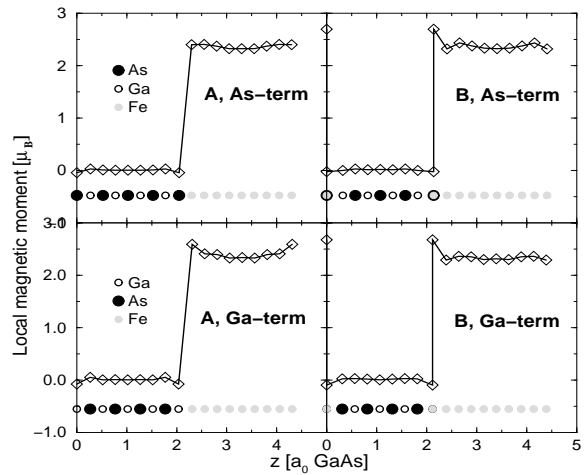}
\caption{ Local magnetic moments plotted as a function of distance $z$
along the direction normal to the interface. 
The upper panels show results for
relaxed As-terminated models A and B, and the lower panels show results for
relaxed Ga-terminated models A and B.
In the intermixed layer in model B,
the higher data point represents the Fe moment 
while the lower data point represents the Ga or As moment. 
\label{fig: mag}}
\end{figure}
Fig. \ref{fig: mag} shows the local magnetic moments across the Fe/GaAs 
interface for all four structurally relaxed models. 
The magnetic moments were calculated from the 
integrated spin-polarized DOS within the Wigner-Seitz spheres. 
There is a small enhancement of the magnetic moment near the Fe 
interface in comparison to the bulk-like moment of $\sim$2.3$\mu_B$
in the central Fe layer.  This is similar what was found in
Ref.[\onlinecite{Vlaic}]. Since the Fe moment at the interface is
sensitive to the Fe-As(Ga) bond length and can be quenched by
reducing the bond length by a few percent,\cite{Mirbt} the relaxation of the
interlayer distances near the interface is important.
In model A the enhancement of the Fe moment at the interface ranges from 
0.1$\mu_B$ for As termination to 0.3$\mu_B$ for Ga termination. 
Model B has a larger enhancement of the magnetic 
moment at the Fe sites located in the intermixed layer, with spin moments 
$\sim0.4\mu_B$ larger than at the bulk-like sites. 
These results, which indicate that Fe is ferromagnetic at the GaAs
interface, are consistent with recent experiments on the Fe/GaAs 
(100)-4$\times$6 interface\cite{bulk_like} where bulk-like spin magnetic moments were 
observed using x-ray magnetic circular dichroism. 
In our calculations, all four structural models also have small induced 
opposite magnetic moments in the first semiconductor layer
with the approximate values of  0.04$\mu_B$ and 0.02$\mu_B$ for
As-terminated models A and B, respectively, and  
0.08$\mu_B$ and 0.09$\mu_B$ for Ga-terminated models A and B, respectively. 

\subsection{Spin-Polarization and Interface States}
%
\begin{figure}[t]
\includegraphics[width=3.0 in,height=2.5 in]{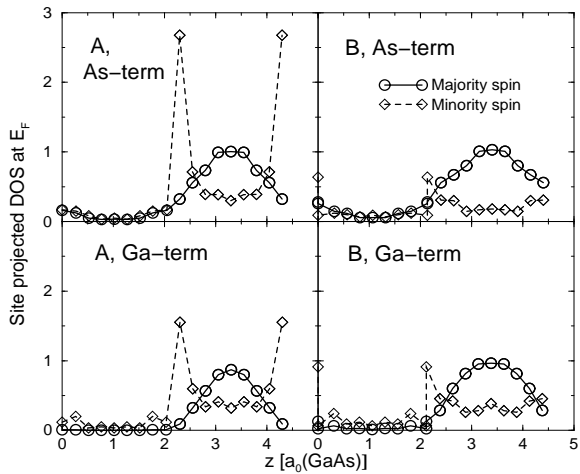}
\caption{Local majority-spin (circles) 
and minority-spin (diamonds) densities of states at the Fermi 
level in different atomic layers. The top panels show results for
relaxed As-terminated models A and B, and the bottom panels show results
for relaxed Ga-terminated models A and B. 
The vertical scale, in units of states/eV/atom, is kept the 
same in all panels to facilitate comparisons. 
\label{fig: SP}}
\end{figure}

Figure \ref{fig: SP} shows the local majority (spin-up) 
and minority (spin-down) densities of states calculated 
at the Fermi energy for the relaxed As- and Ga-terminated models A and B. 
Each point on the curves corresponds to an atomic layer. 
In all four cases the majority-spin DOS is largest at the center 
of the Fe slab and decays monotonically towards and across the interface 
into the GaAs slab.
At the same time the minority-spin DOS exhibits a sharp peak in the vicinity 
of the interface on the Fe side and decreases into the center of the Fe slab. 
In the GaAs slab both majority and minority DOS decay exponentially, 
but the dominance of the minority states is preserved throughout the GaAs slab. 
Hence, close to the interface on the Fe side,  the spin polarization reverses 
sign and peaks because of the large
difference between spin-up and spin-down density of states. 
While model B exhibits such a reversal with relatively modest
differences between spin-up and spin-down DOS, 
model A shows significantly larger values of spin polarization.

\begin{figure}[t]
\includegraphics[height=3.0 in]{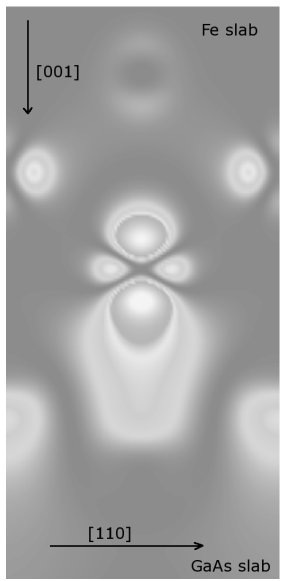}
\caption{Valence charge density of a minority-spin interface state at 
the $\Gamma$-point in the relaxed As-terminated model A supercell.  
The top half of the figure contains Fe layers
and the bottom half contains GaAs layers. 
This state lies about 0.4 eV above the Fermi level.  
\label{fig: surf_state}}
\end{figure}

Similar behavior of the spin-dependent DOS at the Fermi level has
been observed both theoretically and experimentally 
for the free Fe surface. \cite{surf1,surf2} 
A peak in the scanning tunneling spectra of the Fe(001) surface 
was attributed to 
a minority-spin surface band located about 0.3 eV above the Fermi level 
at the $\bar{\Gamma}$ point in the two-dimensional Brillouin zone.
We find analogous behavior at the Fe/GaAs interface, with the 
increase in minority spin DOS at the Fermi level 
at the interface attibutable to $d$ states 
localized on Fe interface sites.  The valence charge
density for one of these interface states at the As-terminated
abrupt interface is shown in Fig. \ref{fig: surf_state}.
This $d_{3z^2-r^2}$-derived  Shockley-like state, located about 0.4 eV 
above the Fermi level 
at the $\Gamma$ point, produces high densities of states within the
GaAs band gap and decays evanescently into the GaAs slab.
At the intermixed interfaces of model B, analogous states
localized on the Fe sites in the intermixed layer are found near
the Fermi level, but because of the stronger influence of the
reduced symmetry of GaAs compared to Fe, some of these interface states
are more clearly a mixture of $d_{3z^2-r^2}$ and $d_{xy}$ character,
with lobes pointing along the $z$ and $x=\pm y$ directions. (The 
$x$ and $y$ directions are along the cubic axes of both the bcc Fe and
zinc-blende GaAs lattices.)

These interface states likely play a role in 
Fermi-level pinning, 
which experimentally manifests as an insensitivity of the SBH 
to the metal workfunction. 
Our calculations suggest that at the defect-free interface, 
the Fermi level is pinned by 
Fe minority-spin interface states, supporting the
metal-induced gap states (MIGS) model \cite{MIGs,MIGs2}
rather than the semiconductor surface state model. \cite{Bardeen} 
This also agrees with recent experiments on the pressure dependence 
of metal/GaAs Schottky barrier heights, \cite{experiment_Jonker2} which 
support the MIGS model for atomically clean Fe/GaAs interfaces. 
A notable difference between the interface states we find and MIGS suggested  
in the original works \cite{MIGs,MIGs2} is that the interface states in our case are 
localized at the interface and are not derived from the metal bulk
states.

To quantify the decay of the spin polarized DOS
into the GaAs, we have performed additional large supercell calculations 
using the relaxed interface geometries, increasing the number of atomic layers 
in the GaAs slab to 17. 
A large number of {\bf k}-points was used in the DOS calculations 
(24$\times$24$\times$4), 
and a Gaussian broadening of states with $\sigma=0.1$ eV was used. 
We fit the DOS at the Fermi level in the GaAs slab 
to the functional form $e^{-2\kappa z}$, where $\kappa$ is a decay constant, 
and $z$ is the distance 
from the interface. The decay constants $\kappa_{\uparrow,\downarrow}$ 
for states with different spin, and corresponding decay lengths 
$l_{\uparrow,\downarrow}$, are listed in Table \ref{tab: decay}. 
\begin{figure}[t]
\includegraphics[width=3.0 in,height=2.70 in]{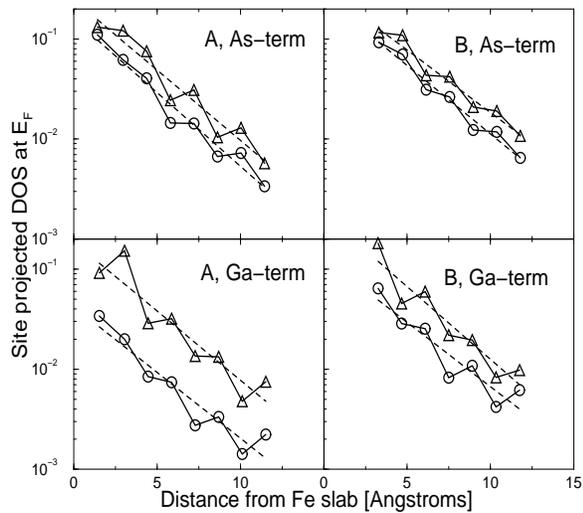}
\caption{The site-projected GaAs DOS at the Fermi level as a function of the 
distance from the Fe slab. The majority-spin DOS are plotted with circles  and
the minority-spin DOS are plotted with triangles.  
The upper panels correspond to As-terminated models A and B, and the
lower panels correspond to Ga-terminated models A and B.   The
DOS are in units of states/eV/atom. 
\label{fig: decay}}
\end{figure}
Semi-logarithmic plots of the majority- and minority-spin DOS at the Fermi 
level in the GaAs slab are shown in Fig. \ref{fig: decay},  
along with the fitted exponentials (dashed lines).
Each point corresponds to an atomic layer, and data corresponding to the 
intermixed layer in model B was omitted since this layer contains both
Fe and semiconductor atoms. 
At the Fermi level the minority-spin states continue to dominate the 
majority-spin states throughout the GaAs slab. There is a significant proximity 
effect in the sense that the {\sl ratio} of the DOS corresponding to the
two spin states does not change significantly. 
In other words, the spin polarization of the states at the Fermi 
level remains virtually constant into the bulk of GaAs, even
as the magnitude of the DOS decays exponentially. 

\begin{table}[h]
\caption{Decay constants $\kappa_{\uparrow,\downarrow}$ (in \AA$^{-1}$) for majority and 
minority states and corresponding decay lengths $l_{\uparrow,\downarrow}$ (in \AA).
\label{tab: decay}}
\begin{ruledtabular}
\begin{tabular}{ccccc}
  & \multicolumn{2}{c}{As-terminated} & \multicolumn{2}{c}{Ga-terminated} \\
\hline
  & $\kappa_{\uparrow}$ & $\kappa_{\downarrow}$ & $\kappa_{\uparrow}$ & $\kappa_{\downarrow}$ \\
\hline
Model A & 0.17 & 0.16 & 0.15 & 0.16 \\
Model B & 0.16 & 0.14 & 0.15 & 0.18 \\
\hline
  & $l_{\uparrow}$ & $l_{\downarrow}$ & $l_{\uparrow}$ & $l_{\downarrow}$ \\
\hline
Model A & 3.0 & 3.1 & 3.3 & 3.1 \\
Model B & 3.2 & 3.5 & 3.4 & 3.0
\end{tabular}
 \end{ruledtabular}
\end{table}
The calculated decay constants $\kappa_{\uparrow,\downarrow}$ listed in Table \ref{tab: decay}
show that there is not a significant difference between the decay of 
spin-up and spin-down states for a given structure geometry. 
Such behavior is expected (assuming that the effective masses of 
spin-up and spin-down electrons do not differ significantly) since
states of the same
energy should have the same decay constants unless the potential barriers 
for the two particles differ. 
The similarities of decay constants show that the Schottky barrier 
heights for both spin-up and spin-down electrons are the same, which 
is reasonable since the GaAs slab is essentially non-magnetic. 
This finding, however, is in 
contrast with results of calculations of the electronic 
properties of Co/Al$_2$O$_3$/Co magnetic tunnel junctions, \cite{Tsymbal} 
where different decay constants for the spin-polarized states at the Fermi 
energy in Al$_2$O$_3$, and therefore different potential barrier heights, 
were obtained for electrons of opposite spins. 
In terms of structure dependence, we find that the interfaces with intermixing 
of metal and semiconductor atoms tend to have longer decay lengths than 
abrupt interfaces, which is consistent with trends in the calculated
Schottky barrier heights.

\begin{figure}[t]
\includegraphics[width=2.5 in]{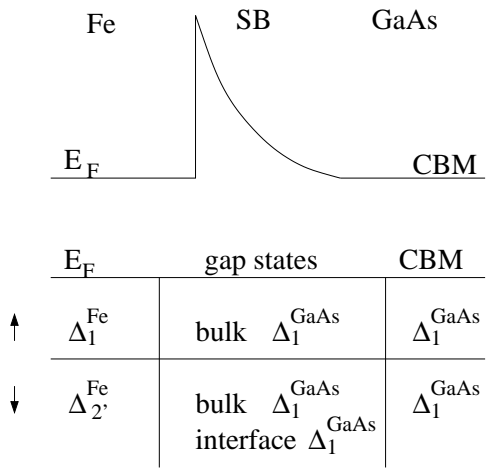}
\caption{Schematic flat-band diagram showing 
that electrons from bulk states near the Fermi level in Fe
tunnel through decaying gap states in the Schottky barrier
into states near the conduction band minimum in GaAs. 
(Note that the width of the barrier region is typically
much larger than the decay length of the gap states.)
In both majority and minority spin channels, there
are bulk-like states in each region that are 
compatible by symmetry.  Only in the minority spin channel are
there interface-derived gap states of the right symmetry near
the Fermi level. 
\label{fig: symm}}
\end{figure}

While detailed transport calculations, using, for 
example, the Landauer or Kubo formalisms,  are needed for  
quantitative predictions regarding spin-polarized current, 
symmetry considerations allow us to make some qualitative arguments 
about the spin-injection process through the interface. 
Electrons tunnel from bulk states near the Fermi level in Fe 
through evanescently decaying gap states in the Schottky barrier region into
states near the conduction band minimum in bulk GaAs (Fig. \ref{fig: symm}). 
The states participating in this transport process must be compatible 
by symmetry. 
In the approximation that transport across the interface is dominated by
carriers with wave vector perpendicular to the interface, we focus on
states with ${\bf k}_{\parallel} \approx 0$ (i.e., along the $\Delta$ 
direction in the cubic Brillioun zone of both GaAs and Fe).
The $1\times1$ interface considered here
has $C_{2v}$ symmetry, and so does the lowest conduction band of
GaAs (indexed as ${\Delta}_1^{GaAs}$).  In the Schottky barrier region,
the most important bulk states in the complex band structure 
are those with the longest decay length.  In GaAs, these states have 
${\Delta}_1^{GaAs}$ symmetry as well.\cite{Mavropoulos}
Therefore, if we consider only bulk-like states, carriers must originate
from states in the bulk Fe bandstructure that are compatible by symmetry with
${\Delta}_1^{GaAs}$ states.  In the majority-spin band structure of Fe,
the $d_{3z^2-r^2}$-derived band, indexed as $\Delta_1^{Fe}$,
crosses the Fermi level  and
satisfies the symmetry requirement.  In the minority-spin band structure,
the $d_{xy}$-derived band, indexed as $\Delta_{2^\prime}^{Fe}$,
crosses the Fermi level and satisfies the symmetry requirement.
Since the $\Delta_1^{Fe}$ band has some $s$ character
as well as $d_{3z^2-r^2}$ character, it is expected that
the coupling of states across the interface will be
stronger in the majority channel. 
Thus, considering only bulk-like states, the majority spin
current would be expected to dominate. In fact, transport calculations
based on the Landauer formalism support this conclusion.\cite{Dederichs2}

However, in addition to bulk-like states, interface states 
may play a role in the tunneling process.  The interface
states in the vicinity of the Fermi level all have minority spin.
Arising from $d_{3z^2-r^2}$ and $d_{xy}$ orbitals on interface
Fe sites, and resonant with the bulk Fe $\Delta_{2^\prime}^{Fe}$ minority
band that crosses $E_F$, these states provide additional symmetry-compatible
gap states for tunneling  of minority spins through the barrier.    Hence
the presence of these interface states close to the Fermi level  could
reduce the spin polarization of the tunneling current, or even reverse
its sign.  In Table  \ref{tab: interface}, 
we list the energy of the symmetry-compatible 
interface states within 0.5 eV of the Fermi level at the $\Gamma$ point 
for different structural models.  These states typically
have dispersions of a few tenths of eV across the Brillouin zone. 
Since the proximity of these interface states to the Fermi level
depends on the atomic structure at the interface, 
we expect the spin polarization of the injected current
to be sensitive to interface structure as well. 
Measurements of the sign of the circular polarization of
electroluminescence from  Fe/GaAs spin LEDs  indicate
injection of minority spins,\cite{Hanbicki}  opposite to what 
is predicted from transport calculations for ideal junctions.\cite{Dederichs2}
The present results indicate that details of the interface structure, such as
the degree of intermixing and relaxation, likely
contribute to this discrepancy.

\begin{table}[tbhf]
\caption{Energy of symmetry-compatible minority-spin interface states 
within 0.5 eV
of the Fermi level.  Energies are in eV and measured relative to
the Fermi level.  
\label{tab: interface}}
 \begin{ruledtabular}
\begin{tabular}{ccccc}
  & \multicolumn{2}{c}{As-terminated} & \multicolumn{2}{c}{Ga-terminated} \\
\hline
  & Ideal & Relaxed & Ideal & Relaxed \\
\hline
Model A & 0.15, 0.16, 0.37 & 0.31, 0.39, 0.39 & N/A & 0.43 \\
Model B & 0.34, 0.41 & 0.16, 0.48  & -0.02 & 0.12, 0.28 \\ 
\end{tabular}
 \end{ruledtabular}
\end{table}

\section{Conclusions}
We have investigated the electronic and magnetic properties of 
Fe/GaAs (001) interfaces using density functional calculations. 
Two structural models were compared: one with an abrupt interface and
one with intermixing of metal and semiconductor atoms.  
Both As- and Ga-terminations were considered,  and interlayer
separations were relaxed.  Due to differences in Fe-As and Fe-Ga
bonding, the  As-terminated structure favors the abrupt  interface
while the Ga-terminated structure favors intermixing at the interface.
Magnetization profiles show bulk-like magnetic moments at the interfaces. 
In all cases, charge is transferred from Fe to GaAs, creating
Schottky barriers that vary in height depending on details of the
interface structure. In general, the SBHs are less sensitive to
interlayer relaxations than to the nature of the interface, with intermixing
of atoms at the interface leading to smaller $n$-type barrier heights. 
In all the structural models considered, the minority-spin 
Fe interface state induces  states of $\Delta_1^{GaAs}$ symmetry
within the semiconductor gap. These states lead 
to a reversal of the sign of the spin polarization of the density of states 
near the interface (compared to the Fe bulk), and 
this spin polarization of the density of states at the Fermi level
persists well into the semiconductor.  These interface-induced gap
states (which also are likely to play a major part in pinning of the Fermi
level) provide an additional channel 
for tunneling of minority spins.  The proximity of these interface
states to the Fermi level, which affects the
magnitude and possibly the sign of the spin polarization of the
tunneling current, varies significantly with interface structure.

\acknowledgments
We would like to acknowledge support from the National Science Foundation,
Grant No. DMR-0210717, and the Office of Naval Research, Grant No. 
N00014-02-1-1046. 
We also thank the National Partnership for Advanced Computational 
Infrastructure for a supercomputing allocation. 



\begin{thebibliography}{99}
\bibitem{WS} The Wigner-Seitz radii were chosen to be 1.355 {\AA} for As, 1.402 {\AA}  for Ga, and 1.302 {\AA} for Fe atoms.
\end{thebibliography}
\end{document}